\documentclass[%
 reprint,
 amsmath,
 nofootinbib,
 amssymb,
 aps]{revtex4-2}

\usepackage{tabularx}
\usepackage{comment}
\usepackage{booktabs}
\usepackage{float}
\usepackage{braket}
\usepackage[unicode=true,bookmarks=false,breaklinks=false,pdfborder={0 0 1},colorlinks=true]{hyperref}
\hypersetup{citecolor=blue,linkcolor=blue,urlcolor=blue}
\usepackage{graphicx}
\usepackage{caption,subcaption} 
\usepackage{dcolumn}
\usepackage{bm}
\usepackage{dsfont}
\def\unit{\mathds{1}}
\usepackage{orcidlink}
\def\unit{\mathds{1}}

\graphicspath{ {./plots/} }

\begin{document}

\preprint{APS/123-QED}

\title{When Bob orbits Alice: entanglement harvesting in circular motion}

\author{F. Sobrero \orcidlink{0009-0009-6497-6612}}
 \email{felipesobrero@cbpf.br}
\affiliation{Centro Brasileiro de Pesquisas F\'isicas, Rua Xavier Sigaud, 150 - Urca, Rio de Janeiro, RJ, Brazil.}
\author{M. S. Soares \orcidlink{0000-0001-5000-952X}}%
 \email{matheus.soares@cbpf.br}
\affiliation{Centro Brasileiro de Pesquisas F\'isicas,  Rua Xavier Sigaud, 150 - Urca, Rio de Janeiro, RJ, Brazil.}
\author{N. F. Svaiter \orcidlink{0000-0001-8830-6925}}%
 \email{nfuxsvai@cbpf.br}
\affiliation{Centro Brasileiro de Pesquisas F\'isicas,  Rua Xavier Sigaud, 150 - Urca, Rio de Janeiro, RJ, Brazil.}

\begin{abstract}
    We study radiative processes of two qubits coupled to a massless scalar field prepared in the Minkowski vacuum state. The analyze the effects of vacuum fluctuations in the generation of qubits' entangled states is performed. We assume one of the qubits is at rest in an inertial frame while the other comoves with a uniformly rotating frame, $i.e.$, undergoing circular motion. We investigate how the entanglement harvesting phenomenon depends on the radius and angular velocity of the non-inertial qubit. We compute the concurrence and mutual information to identify the set of circular motion parameters that maximizes entanglement generation.

\end{abstract}

\maketitle

\section{Introduction}\label{sec: introduction}

The interplay between quantum information and relativity, often called relativistic quantum information, addresses how quantum resources (entanglement, coherence, quantum correlations) behave when systems move non-inertially in Minkowski spacetime or are defined in a curved spacetime. This topic not only probes that the definition of particles associated to a quantum field are observer-dependent, but also information transfers between different points in the framework of quantum field theory. These subjects have operational implications for quantum technologies in relativistic settings, $e.g.$, satellite quantum communications and experiments that probe fundamental quantum field theory effects. The core of the studies in relativistic quantum information is motivated by two closely related predictions from quantum field theory formulated in non-inertial frames and curve spacetimes: the \emph{Unruh--Davies effect}: an uniformly accelerated observer perceives the Minkowski vacuum state as a thermal bath \cite{Davies:1974th, Unruh:1976db}; and the \emph{Hawking effect} which is related to the thermal radiation of a black hole \cite{hawking75}. These two results exemplifies how particle content and the effect of vacuum fluctuations depend on the observer’s trajectory and horizons \cite{dewitt75, sciama1981quantum, Bire82, fulling89, ford2002}.

There are many different approaches in the literature to study, in relativistic settings, information issues in Minkowski spacetime. The two most used approaches are based on: (a) working with \emph{mode decompositions and Bogoliubov transformations} between field basis constructed by inertial and non-inertial observer \cite{takagi86} and (b) quantum systems as localized \emph{two-level systems or qubits} that couple to the quantum field through a time-dependent switching (coupling) function.

The first one is useful to quantify how global field positive and negative frequency modes become mixed for different observers worldlines and how mode entanglement is redistributed. The second one consider the \textit{qubit-quantum field coupling}, commonly the Unruh--DeWitt model or localized two-level systems  \cite{zhou2013boundary,martin2016spacetime,menezes2016radiative,arias2016boundary,ZhangYu2020,bozanic2023correlation}. However, the atom-field interactions can be modelled in several ways for completely different contexts, see Refs. \cite{hsiang2024atom,hsiang2025atom,hsiang2026atom}. In the first scenario, one can perform different investigations on the qubits' dynamics. For example, after tracing out the field, the qubits’ reduced density matrix encodes the effective open-system dynamics and allows  direct calculation of qubit excitation, decoherence and entanglement dynamics \cite{svaiter1992, soares22, soares25}. It can be generalized to curved spacetimes with a very similar approach \cite{tjoa2023}. Time-evolution techniques for multiple qubits, $e.g.$, master equations, are widely used to obtain both analytic and numerical results. These complementary tools have clarified phenomena such as entanglement degradation for uniformly accelerated qubit, entanglement harvesting from the vacuum, and the role of switching functions and detector spatial profile in physically realistic scenarios~\cite{Bruschi2010,Bruschi2012,MartnMartnez2015}.

\begin{figure}[ht!]
\centering
\includegraphics[scale=0.3]{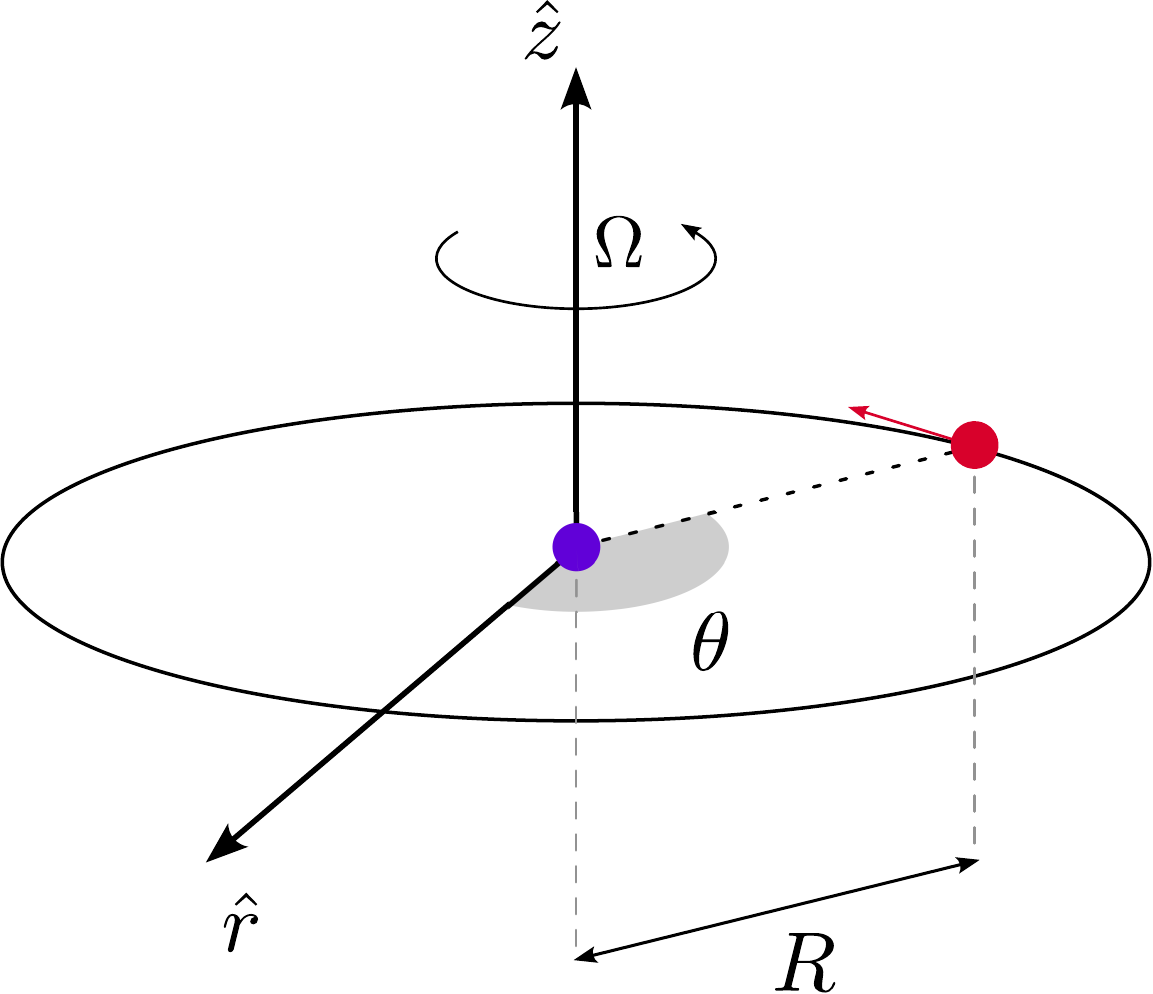}
\caption{Diagram representing two atoms: the atom 1 and the atom 2. Alice is at the center of the coordinate system at a distance $R$ of Bob, which rotates in a circular motion around Alice.}
\label{fig: entangled atoms}
\end{figure}

In this work we focus on a setting that highlights trajectory-dependent entanglement generation: \emph{two localized two-level systems (Alice and Bob's qubits)} initially prepared in its ground states has a non-zero probability to make a transition to the maximally entanglement Bell's state, where Alice follows an inertial worldline and Bob undergoes uniform circular motion with radius $R$ and angular velocity $\Omega$ as illustrated in Fig. \ref{fig: entangled atoms}. Both qubits interact locally with a quantum scalar field and the full closed evolution is given by the von Neumann's equation for the combined qubits and field state. Also, the qubits’ dynamics of interest are obtained by tracing out the field's degrees of freedom. Our main goal is to characterize how the bipartite entanglement of the qubits reduced density matrix increases as a function of Bob’s orbital radius (and related kinematic parameters). Operationally, we quantify entanglement generation through concurrence~\cite{ZhangYu2020,Pan2024}.

The issue of radiative processes for two-level systems in a non-inertial rotating frame has been addressed in several works \cite{denardo1978quantum, letaw80, bell1983electrons, bell1987unruh, doukas2013unruh}. In the literature, the discrepancy between the physical interpretation derived from Bogoliubov coefficients, relating modes defined by rotating and inertial observers, and the qubits’ response function is discussed. Since the coefficient $\beta_{ij}$ vanish, the Minkowski vacuum state and the one defined for rotating observers are unitarily equivalent. Nevertheless, a qubit in a rotating frame interacting with a scalar field in the Minkowski vacuum state exhibits a non-zero response function for excitations. Regarding the two-level system as a detector, a solution for this apparent inconsistent situation was presented by Davies et al. \cite{davies1996detecting}. Another way to avoid such contradiction is to use the rotating wave approximation as a necessary condition to define a detector \cite{glauber1963I, glauber1963II}. In Refs. \cite{soares21, sobrero2025}, the operational definition of a measurement device was examined, where the Unruh-DeWitt and Glauber models of detectors were discussed.

Beyond this inconsistency between the response function and Bogoliubov approaches, a deeper question arises: how can rotation be incorporated into a relativistic framework? This problem has attracted considerable attention from physicists such as Ehrenfest, Born, Planck, Kaluza, Einstein, and others \cite{ehrenfest1909gleichformige, born1909theorie, planck1910gleichformige, einstein1910ehrenfestschen}. Landau and Lifshitz \cite{landau1975classical} proposed a transformation law between cylindrical coordinates adapted to inertial and rotating frames, valid only for $r < c/\omega$. To extend this system to arbitrary radius and avoid tangential velocities exceeding $c$ for $r > c/\omega$, Fraklin \cite{franklin22} and later Trocheries and Takeno \cite{takeno1952relativistic} have introduced a coordinate system where $v/c = \tanh(\omega r / c)$, which approaches $c$ asymptotically. The implications of this transformation for field theory are discussed in \cite{de1999rotating, de2000rotatingdetector, de2001rotating}. However, this choice fails to reproduce experimental results such as the Sagnac effect \cite{sagnac1913ether}, where an interferometer on a rotating disk measures the phase shift between two coherent light beams traveling in opposite directions. An alternative proposal by Gr{\o}n \cite{gro1975relativistic, gron1977rotating} successfully reproduces the Sagnac effect but introduces a discontinuity in the time coordinate for closed circuits around the origin. A different approach to the kinematics of rotating frames was later developed by Klauber \cite{klauber1998new}. See also Ref. \cite{rizzi2004relativity}. Until now, there is no consensus on how to characterize the relativistic rotating disk in the literature.

Based on previous studies of two-level systems in circular motion and of initial entanglement between noninertial parties, we expect several characteristic features: (i) entanglement typically degrades faster when Bob’s qubit feels stronger effective vacuum fluctuations (higher radial acceleration or higher angular velocity at fixed radius), (ii) non-thermal signatures may appear for circular motion, while uniformly linearly accelerated qubits have a thermal response in the idealized long-time limit, rotating qubits show parameter regimes where the effective temperature picture fails and the response depends sensitively on the radius and switching duration, (iii) switching functions and finite interaction times can produce transient revivals or oscillatory behavior in concurrence, and (iv) the \emph{massive field} and geometry (e.g., presence of mirrors or cavity boundaries) can qualitatively change the rate and pattern of entanglement loss. Quantifying these effects for an inertial Alice and a rotating Bob will allow us to identify parameter regimes where relativistic 
motion strongly restricts usable entanglement and regimes where entanglement is comparatively robust.

This paper is organized as follows. In Sec.  \ref{sec:quantization} we introduce the quantization of a massless scalar field in rotating frames and define the positive frequency two-point Wightman functions. In Sec.  \ref{sec: unruh-dewitt detectors} we present the two-level systems model and the interaction Hamiltonian with switching functions. Next, we review the procedure to obtain the two-atom reduced density matrix by solving the von 
Neumann equation and tracing out the field's degrees of freedom and finally define the concurrence. In Sec. \ref{sec:results} we compute the transition probabilities and discuss the entanglement harvesting as a function of Bob’s radius and qubit's energy gap. Conclusions are given in Sec. \ref{sec:conclusion}. In this work we consider $c = \hbar = 1$.

\section{Canonical quantization}\label{sec:quantization}

Let us start by considering a scalar field $\phi(t,\mathbf{x})$ defined for a four-dimensional Minkowski spacetime, satisfying the massless Klein-Gordon equation
	\begin{equation}\label{eq: KG}
		\Box \phi = 0,
	\end{equation}
where $\Box = \eta_{\mu\nu}\partial^\mu \partial^\nu$. We wish to obtain a set of solution of the above equation by considering a coordinate system adapted to a \textit{uniformly rotating frame}. This can be done by performing a coordinate change of the above equation using a cylindrical coordinate system $x'^{\mu} = \{t',r',\theta',z'\}$.

The Klein-Gordon equation, in such coordinate system is given by
	\begin{equation}
\Bigg(\partial^2_{t'} - \partial^2_{r'} - \frac{1}{r'}\partial_{r'} - \frac{1}{r'^2}\partial^2_{\theta'} - \partial^2_{z'} \Bigg) \phi = 0
	\end{equation}
whose one set of solutions is given by
\begin{equation}\label{eq: modes}
u_{kmq}(x') = \frac{1}{2\pi\sqrt{2\omega_{kq}}} e^{-i \omega_{kq}t' + i kz' + i m \theta'}J_m(q r'),
\end{equation}
where $\omega^2_{k,q} = k^2 + q^2$ and $J_\ell(x)$ is a Bessel function. The field must be integrated over $k$ and $q$ and summed over $m$. The modes defined in Eq. \eqref{eq: modes} are said to be positive frequency with respect to $t'$, being eigenfunctions of the operator $\partial_{t'}$:
\begin{equation}
\partial_{t'} u_{kmq}(x') = -i\omega_{kq} u_{kmq}(x'), ~\omega_{kq} > 0.
\end{equation}
By defining the scalar product
\begin{equation}
(\phi_1,\phi_2) = -i \int d^3\mathbf{x} \Big(\phi_1 \partial_t \phi^*_2 - (\partial_t \phi_1) \phi^*_2 \Big),
\end{equation}
where here $t$ is a spacelike hyperplane of simultaneity, the modes $u_\mathbf{k}$ from Eq. \eqref{eq: modes} are orthogonal and normalized, such that
\begin{equation}
(u_{kmq},u_{k'm'q'}) = \delta_{m m'}\delta(k-k')\delta(q-q').
\end{equation}
Once we obtain the complete set of modes solutions of the field equations in the cylindrical coordinates one can expand the fields using these modes. In the canonical quantization framework a quantized scalar field in Minkowski spacetime can be expressed as
	\begin{eqnarray}
    &&\phi(x') 
    \label{eq: inertial field}\\
    &&= \sum_{\substack{m=-\infty \\ \omega_{kq} > 0}}^{\infty}\int_{-\infty}^{\infty} q dq dk \Big( a_{kmq}u_{kmq}(x') + a^\dagger_{kmq}u^*_{kmq}(x') \Big), \nonumber
	\end{eqnarray}
where the operators $a_{kmq}(a_{kmq}^\dagger)$, the annihilation (creation) operators of quanta of the field, obey the commutation relation:
	\begin{equation}\label{eq: commutation relation}
[a_{kmq},a^\dagger_{k'm'q'}] = \frac{\omega_{kq}}{q|\omega_{kq}|}\delta_{\omega \omega'}\delta_{m m'}\delta(k-k')\delta(q-q'),
	\end{equation}
while $[a_{kmq},a_{k'm'q'}] = [a^\dagger_{kmq},a^\dagger_{k'm'q'}] = 0$. These operators allows us to define a Fock space with a vacuum state $\ket{0_M}$, defined by
	\begin{equation}
    a_{kmq}\ket{0_M} = 0, ~\forall ~k,m,q.\label{eq:vacuumstate}
	\end{equation}

The above discussion was restricted to a cylindrical coordinate system adapted to an inertial frame. We are interested in the study of the canonical quantization of the scalar field implemented by observers at rest in a frame comoving with a disk in uniformly circular motion. In this case, we must add a constant angular velocity $\Omega$ to the discussion. A transformation to a rotating coordinate is described by
\begin{equation}
t' = t, ~r' = r, ~\theta' = \theta + \Omega t, ~z' = z,
\end{equation}
and provides the following line element
	\begin{equation}
		ds^2 = (1-\Omega^2 r^2)dt^2 -2\Omega r^2 d\theta dt 		- r^2 d\theta^2 - dr^2 - dz^2.
	\end{equation}
This metric is stationary but not static, i.e., $\partial_t$ is not orthogonal to a family of spacelike hypersurfaces. The adaptation of Killing vector to an observer orbiting radius $R$ is completed by replacing $\partial_t$ by $\gamma \partial_t$ where $\gamma = (1 - \Omega^2 R^2)^{-1/2}$, which is normalized along that observer's world line.

The Klein-Gordon equation, given by Eq. \eqref{eq: KG} becomes,
\begin{equation}
\Bigg( \Big( \partial_t - \Omega \partial_\theta\Big)^2 - \partial^2_r - \frac{1}{r}\partial_r - \frac{1}{r^2}\partial^2_\theta - \partial^2_z \Bigg)\phi = 0,
\end{equation}
therefore, as shown by Letaw and Pfautsch in Ref. \cite{letaw80}, the mode functions for the rotating coordinate system is identical in appearance to the ordinary Minkowski mode function transformed to rotating coordinates:
\begin{equation}\label{eq: rotating modes}
v_{kmq}(x) = \frac{1}{2\pi\sqrt{2\omega_{kq}}} e^{-i (\omega_{kq} - m \Omega)t + i kz + i m \theta}J_m(q r).
\end{equation}
The field operator $\phi$ can also be expanded in terms of the field modes defined in Eq. \eqref{eq: rotating modes} as
\begin{eqnarray}
    &&\phi(x) \label{eq: rotating field} \\
    &&= \sum_{\substack{m=-\infty \\ \omega_{kq} - m\Omega > 0}}^{\infty}\int_{-\infty}^{\infty} q dq dk \Big( b_{kmq}v_{kmq}(x) + b^\dagger_{kmq}v^*_{kmq}(x) \Big), \nonumber
\end{eqnarray}
then the commutation relations from $b$ and $b^\dagger$ are exactly the same as the one given by $a$ and $a^\dagger$ in Eq. \eqref{eq: commutation relation}
\begin{equation}\label{eq: rotating commutation relation}
[b_{kmq},b^\dagger_{k'm'q'}] = \frac{\omega_{kq}}{q|\omega_{kq}|}\delta_{\omega \omega'}\delta_{m m'}\delta(k-k')\delta(q-q'),
\end{equation}
while $[b_{kmq},b_{k'm'q'}] = [b^\dagger_{kmq},b^\dagger_{k'm'q'}] = 0$. These operators allow us to define a Fock space with a rotating vacuum state, $\ket{0_R}$, defined by
\begin{equation}\label{eq: rotating vacuum state}
    b_{kmq}\ket{0_R} = 0, ~\forall ~k,m,q.
\end{equation}

Comparing the field decomposition of Eq. \eqref{eq: rotating field} with the one defined in Eq. \eqref{eq: inertial field}, we can write the following relations
	\begin{equation}\label{eq: annihilation operator}
b_{kmq} = \left\{ \begin{array}{lc}
a_{kmq}, & \omega_{kq} > 0, \\
(-1)^m a^{\dagger}_{-k m -q}, & \omega_{kq} <0,
\end{array} \right.
	\end{equation}
\begin{equation}\label{eq: creation operator}
b^\dagger_{kmq} = \left\{ \begin{array}{lc}
a^\dagger_{kmq}, & \omega_{kq} > 0, \\
(-1)^m a_{-k m -q}, & \omega_{kq} <0.
\end{array} \right.
	\end{equation}
By analyzing the above equation, one can say that the effect of a rotating observer is to replace the field modes for $\omega_{kq} > m\Omega$ with the time reversed solutions for $\omega_{kq} < 0$ \citep{letaw80}. We can now define the number operator in the rotating frame by
\begin{equation}
N_R = \sum_{\substack{m=-\infty \\ \omega_{kq} - m\Omega > 0}}^{\infty}\int_{-\infty}^{\infty} q dq dk \Big( \Theta(\omega) b^\dagger b + \Theta(-\omega) b b^\dagger \Big).
\end{equation}
Using Eqs. \eqref{eq: annihilation operator} and \eqref{eq: creation operator} to rewrite this operator in terms creation and annihilation operators associated to Minkowski vacuum, we obtain that $N_R = N$, where $N$ is the number operator in the inertial frame and we have
	\begin{equation}
\bra{0_M} N_R \ket{0_M} = 0.
	\end{equation}
We can interpret the above equation by saying that there is no excitations associated to Eq. \eqref{eq: rotating field} in the Minkowski vacuum state. This result was first obtained in Ref. \cite{denardo1978quantum}. However, Letaw and Pfautsch \cite{letaw80} have shown that the power spectrum for the rotating observer is not equal to the inertial case. $i.e.$, a qubit in a rotating frame could be excited even if the qubit interact with the field defined in the vacuum state defined by Eq. \eqref{eq: rotating vacuum state}.


The two-point positive frequency Wightman function, $W(x;x') \equiv \bra{0} \phi(t,\mathbf{x}) \phi(t',\mathbf{x}') \ket{0}$, along the worldline of a generic observer in Cartesian coordinates is given by the following expression:
\begin{equation}\label{eq: wightman function}
W(x;x') = \frac{1}{4\pi^2} \int_{0}^{\infty} d\omega ~e^{-i\omega(t - t')} \frac{\sin \omega|\mathbf{x} - \mathbf{x}'|}{|\mathbf{x} - \mathbf{x}'|},
\end{equation}
to avoid misunderstanding, from now on $x$ and $x'$ are related to different points of the spacetime defined by inertial and/or rotating observers.

We will consider two observers in the setup illustrated in Fig. \ref{fig: entangled atoms}. One, Alice, an inertial observer defined in the worldline:
\begin{equation}\label{eq: static observer}
t_A = \tau, ~x_A = y_A = z_A = 0,
\end{equation}
and Bob, a observer at rest in a frame in uniform circular motion with radius $R$ and constant angular velocity $\Omega$. Bob's worldline can be written as
\begin{equation}\label{eq: rotating observer}
t_B = \gamma \tau, ~x_B = R \cos(\Omega t_2), ~y_B = R \sin(\Omega t_2), ~z_B = 0,
\end{equation}
where $\gamma = (1 - \Omega^2R^2)^{-1/2}$. The two-point correlation functions for the inertial observer, the rotating observer and the crossed case can be respectively written as
\begin{align}
&W(x_A;x_A') = \frac{1}{4\pi^2}\lim_{\Delta x \to 0} \int_{0}^{\infty} d\omega ~e^{-i\omega\Delta \tau} \frac{\sin \omega \Delta x}{\Delta x}.\ \label{eq: inertial wigthman}\\
&W(x_A;x_B') = \frac{1}{4\pi^2} \int_{0}^{\infty} d\omega ~e^{-i\omega(\tau - \gamma \tau')} \frac{\sin \omega R}{R}, \label{eq: crossed wigthman 12}\\
&W(x_B;x_B') = \frac{1}{4\pi^2} \int_{0}^{\infty} d\omega ~e^{-i\gamma \omega\Delta \tau} \frac{\sin\left( 2\omega R \left|\sin\frac{\gamma \Omega}{2} \Delta \tau \right|\right)}{2R \left|\sin\frac{\gamma \Omega}{2} \Delta \tau \right|} \label{eq: rotating wigthman}
\end{align}
It is important to note that $W(x_A,x'_B)$ cannot be written as a function of $\Delta \tau$. This can be done only for the regime of non-relativistic velocities, i.e., $\Omega R \ll 1$.

\section{Entanglement dynamic of two qubits}\label{sec: unruh-dewitt detectors}

Now we consider a pair of qubits interacting with a massless scalar field via the following Hamiltonian
\begin{equation}\label{eq:full_hamiltonian}
H(t) = \frac{d\tau_A}{dt} H_A(t) + \frac{d\tau_B}{dt} H_B(t),
\end{equation}
where $t$ is the time parameter we use to describe the evolution of the system and $\tau_i$ is the proper time for each qubit $i = A,B$. The local interaction between each two-level system and the field is
\begin{equation}
H_i(\tau_i) = \lambda \chi(\tau_i) \sigma^{x}_i(\tau_i)\otimes \phi\big(x_i (\tau_i)\big), ~i=A,B.
\end{equation}
where $\lambda$ is the coupling constant which we assume to be small. The $x^\mu_i(\tau)$ are the timelike trajectories of each qubit parametrized by their own proper times. 

The qubit labelled by the letter $A$ is at rest in an inertial frame of reference, the Alice's worldline defined by Eq. \eqref{eq: static observer}. The second quibt, letter $B$, is at rest in a frame in a uniformly circular motion, the Bob's worldline defined in Eq. \eqref{eq: rotating observer}. For simplicity, we are considering identical two-level systems with the same Gaussian switching functions assumed to have no time delay \cite{tjoa2023} 
\begin{equation}\label{eq: switching functions}
\chi(\tau) = e^{-\frac{\tau^2}{\sigma^2}},
\end{equation}
where $\sigma$ prescribes the duration of interaction. The monopole operator $\sigma^{x}(\tau)$ is written as
\begin{equation}
\sigma^{x}_i(\tau) = \sigma^{+}_i e^{i E \tau} +\sigma^{-}_i e^{-i E \tau},
\end{equation}
where $\sigma^{\pm}_i$ are the raising and lowering operator for each two-level system defined by the sub-index $i$.

The qubit-field interaction for a given initial state $\rho_0$ is implemented by unitary time evolution $\rho = U \rho_0 U^\dagger$, where the time evolution operator $U$ is given by the time-ordered exponential
\begin{equation}
U = \mathcal{T} ~\exp\left[-i \int dt ~H(t) \right],\label{eqq:evolutionoperator}
\end{equation}
where $H(t)$ is given by Eq. \eqref{eq:full_hamiltonian} and $\mathcal{T}$ denotes the time ordering operator. By assuming a weak coupling regime, we can perform a series expansion
\begin{equation}\label{eq: dyson expansion}
U = \unit + U^{(1)} + U^{(2)} + \mathcal{O}(\lambda^3),
\end{equation}
whose first two terms, using Eq. \eqref{eqq:evolutionoperator}, are given by
\begin{eqnarray}
U^{(1)} &=& -i \int_{-\infty}^{\infty}d\tau  H(\tau), \\
U^{(2)} &=& - \int_{-\infty}^{\infty}d\tau \int_{-\infty}^{\tau}d\tau'  H(\tau) H(\tau'),
\end{eqnarray}
where $U^{(j)}$ is of order $\lambda^j$. The above expansion allows us to write the density matrix as
\begin{equation}
\rho = \rho_0 + \rho^{(1)} + \rho^{(2)} + \mathcal{O}(\lambda^3),
\end{equation}
where $\rho^{(j)}$ is also of order $\lambda^j$ and are given by
\begin{eqnarray}
\rho^{(1)} &=& U^{(1)}\rho_0 + \rho_0 U^{(1)\dagger}, \label{eq: rho 1}\\
\rho^{(2)} &=& U^{(1)}\rho_0 U^{(1)\dagger} + U^{(2)}\rho_0  + \rho_0 U^{(2)\dagger}. \label{eq: rho 2}
\end{eqnarray}

In order to quantify the entanglement between the two qubits, one must find the joint reduced density matrix of the Alice and Bob's subsystem by tracing out the field's degrees of freedom
\begin{equation}\label{eq:rho_subsystem}
\rho_{AB} = \text{tr}_\phi \left( U \rho_0 U^\dagger\right).
\end{equation}
With the above reduced density matrix we have everything to start the discussion about the entanglement formation. In the next topic we discuss the initial setup of the two qubits and investigate the matrix elements of Eq. \eqref{eq:rho_subsystem}.

\subsection*{Entanglement harvesting protocol}\label{sec: entanglement haversting protocol}

Our objective is to investigate the entanglement harvesting phenomena. To do so, we consider the two qubits prepared in the ground state and the field prepared in the Minkowski vacuum state. This initial state configuration is represented by
\begin{equation}
\rho_0 = \ket{g_A g_B}\bra{g_A g_B}\otimes\ket{0_M}\bra{0_M},
\end{equation}
where, in the right term of the above tensor product, $\ket{0_{M}}$ is the field's vacuum state defined by Eq. \eqref{eq:vacuumstate}. The matrix representation of $\rho_{AB}$ to leading order reads
	\begin{equation}\label{eq: density matrix EH}
\rho_{AB} = \left(\begin{array}{cccc}
1 - \mathcal{L}_+ & 0 & 0 & \mathcal{M}^* \\
0 & \mathcal{L}_{BB} & \mathcal{L}_{AB} & 0 \\
0 & \mathcal{L}_{AB}^* & \mathcal{L}_{AA} & 0 \\
\mathcal{M} & 0 & 0 & 0
\end{array} \right) + \mathcal{O}(\lambda^4),
	\end{equation}
where the basis $\ket{g_Ag_B}, ~\ket{g_Ae_B}, ~ \ket{e_Ag_B}, ~\ket{e_Ae_B}$ has been used and $\mathcal{L}_\pm = \mathcal{L}_{AA} \pm \mathcal{L}_{BB}$. The corresponding elements in $\rho_{AB}$ are given by
\begin{eqnarray}
\mathcal{L}_{ij} &=& \lambda^2 \int d\tau d\tau' \chi(\tau) \chi(\tau') e^{-i E(\tau - \tau')} W\big( x_i,x_j'\big), \label{eq: transition probability} \\
\mathcal{M} &=& -\lambda^2 \int d\tau \int_{-\infty}^{\tau/\gamma}d\tau' \chi(\tau)\chi(\tau') e^{i E(\tau + \tau')} W(x_1,x'_2) \nonumber \\
&&-\lambda^2 \int d\tau' \int_{-\infty}^{\gamma \tau'}d\tau \chi(\tau)\chi(\tau') e^{i E(\tau + \tau')} W(x_2,x_1) \nonumber \\
\label{eq: non-local term}
\end{eqnarray}
where $i,j = \{A,B\}$ and $W(x,x')$ is the positive frequency Wightman function given by Eq. \eqref{eq: wightman function}. The quantities $\mathcal{L}_{AA}$ and $\mathcal{L}_{BB}$ are the transition probabilities of Alice and Bob, respectively, whereas $\mathcal{M}$ and $\mathcal{L}_{AB}$ correspond to `non-local' terms that depend on both trajectories simultaneously \cite{tjoa2023, lopez2025}.

To simplify the computation of the elements of reduced matrix, we define the coordinates given by
\begin{equation}\label{eq: new variables}
    \eta = \frac{\tau + \tau'}{2}, ~\xi = \tau - \tau',
\end{equation}
which allows us to rewritten $d\tau d\tau' = d\eta d\xi$ and also the product of switching functions as $\chi(\tau)\chi(\tau') = e^{-\frac{2\eta^2}{\sigma^2}} e^{-\frac{\xi^2}{2\sigma^2}}$. We also define the dimensionless quantities pondered by the time of interaction $\sigma$
\begin{equation}\label{eq: parameters}
    \varepsilon = E \sigma, ~r = R/\sigma, ~\omega = \Omega \sigma.
\end{equation}
The explicit details of the calculations of $\mathcal{L}_{ij}$ and $\mathcal{M}$ defined by Eqs. \eqref{eq: transition probability} and \eqref{eq: non-local term} can be seen in Appendix \ref{app: A} and \ref{app:B}. For $i = j = A$, we have
\begin{equation}\label{eq: transition probability 11}
    \mathcal{L}_{AA} = \frac{\lambda^2}{4\pi} \left[ \exp\left(-\frac{\varepsilon^2}{2} \right) - \sqrt{\frac{\pi}{2}} \varepsilon ~ \text{Erfc}\left(\frac{\varepsilon}{\sqrt{2}} \right) \right],
\end{equation}
where $\text{Erfc}(z) = 1 - \text{Erf}(z)$. For the case where $i = A$ and $j = B$, we can use $\gamma_\pm = \gamma \pm 1$ and define
\begin{equation}
    \Sigma^2 = \gamma_+^2 + \gamma_-^2,  \quad j_\pm = \frac{\gamma_+ \varepsilon  \pm 2ir}{\sqrt{2} ~\Sigma}. \label{eq: sum of gammas}
\end{equation}
\noindent Therefore, one can write
\begin{align}\label{eq: transition probability 12}
    \mathcal{L}_{AB} = &-\frac{i}{4\sqrt{2\pi}} \frac{\lambda^2}{\Sigma r} e^{-\frac{\varepsilon^2}{2}}\left[ e^{-j_+^2} \text{Erfc}(j_-) - e^{-j_-^2} \text{Erfc}(j_+) \right].
\end{align}
\begin{widetext}

Considering $i = j = B$, we have obtained
\begin{equation}\label{eq: transition probability 22}
    \mathcal{L}_{BB} = \mathcal{L}_{AA} +  \frac{\lambda^2}{4\pi^2} \sqrt{\frac{\pi}{2}}~ \gamma v^2\omega \int_{0}^{\infty} dz ~e^{-2\left(\frac{z}{\gamma \omega}\right)^2} \cos\left(\frac{2\varepsilon z}{\gamma \omega} \right)\frac{z^2 -\sin^2 z}{z^2\left(z^2 - v^2 \sin^2 z \right)},
\end{equation}
where $v = \omega r$. Finally, the term $\mathcal{M}$ was also obtained and has the following expression
\begin{eqnarray}
    \mathcal{M}
    &=&
    -\frac{\sigma^2}{8\pi}
    e^{-\frac{\varepsilon^2}{2}}
    \int_{0}^{\infty}\! dk\, \frac{\sin(k r)}{r}
    e^{-\frac{1}{8}\Sigma^2 k^2} \label{eq: non-local term M} \\
    &&\times \left\{
        2\cosh \left( \frac{\gamma_-}{2} \varepsilon k \right)
        + e^{-\frac{\gamma_-}{2}\varepsilon k} \mathrm{Erf}\!\left[
        \frac{i}{\sqrt{2}}\left( \frac{k \Sigma}{2} + \frac{\varepsilon \gamma_-}{\Sigma} \right)
    \right] + e^{\frac{\gamma_-}{2}\varepsilon k} \mathrm{Erf}\!\left[
        \frac{i}{\sqrt{2}}\left( \frac{k \Sigma}{2} - \frac{\varepsilon \gamma_-}{\Sigma} \right)
    \right]
    \right\}. \nonumber
\end{eqnarray}
\end{widetext}
%

In order to measure the amount of entanglement between the two atoms, we can use the concurrence $C_{AB}$ \cite{wootters1998}. For the time-evolved density matrix in our scenario, the concurrence has the form \cite{tanas2004entangling,martin2016spacetime}
\begin{equation}\label{eq: concurrence}
C_{AB} = 2 \text{max} \left\{ 0,|\mathcal{M}| - \sqrt{\mathcal{L}_{AA} \mathcal{L}_{BB}}\right\} + \mathcal{O}(\lambda^4).
\end{equation}
A naive inspection of the concurrence could lead to the incorrect conclusion that it would be given by an additional condition
: $2|\mathcal{L}_{12}| > 0$. However, this term is not the only contribution at order $\mathcal{O}(\lambda^2)$. In fact, the concurrence arises from the expression $2\left(|\rho_{12}| - \sqrt{\rho_{33}\rho_{44}}\right)$ as can be seen in Eq.~(10) of Ref.~\cite{tanas2004entangling}. Although the matrix element $\rho_{44}$ is nonzero only at order $\mathcal{O}(\lambda^4)$, its contribution enters the concurrence through the square root, yielding an effective $\mathcal{O}(\lambda^2)$ term. This contribution combines with $|\rho_{12}|$ in such a way that the resulting expression is always negative, as discussed in detail in Ref.~\cite{martin2016spacetime}.

The concurrence is zero if and only if the state $\rho_{AB}$ is separable. The mutual information, otherwise, quantifies the amount of total correlation, both classical and quantum \cite{bozanic2023correlation}. For two qubits, mutual information $\mathcal{I}_{AB}$ up to second order in $\lambda$ is \cite{pozas2015harvesting}
\begin{eqnarray}
\mathcal{I}_{AB} &=& \overline{\mathcal{L}}_+ \log \overline{\mathcal{L}}_+ + \overline{\mathcal{L}}_- \log \overline{\mathcal{L}}_- \label{eq: mutual information}\\
&& - \mathcal{L}_{11} \log \mathcal{L}_{11} -  \mathcal{L}_{22} \log \mathcal{L}_{22} + \mathcal{O}(\lambda^4),\nonumber
\end{eqnarray}
where
\begin{equation}
\overline{\mathcal{L}}_\pm = \frac{1}{2}\left( \mathcal{L}_+ \pm \sqrt{\mathcal{L}_-^2 + 4|\mathcal{L}_{12}|^2} \right),
\end{equation}
in which $\mathcal{L}_\pm = \mathcal{L}_{11} \pm \mathcal{L}_{22}$. As discussed in \cite{bozanic2023correlation}, while the concurrence given by Eq. \eqref{eq: concurrence} vanishes when $\sqrt{\mathcal{L}_{11}\mathcal{L}_{22}}$ exceeds the `non-local' element $|\mathcal{M}|$, the mutual information, given by Eq. \eqref{eq: mutual information}, becomes zero when $ |\mathcal{L}_{12}| = 0$. But, if $\mathcal{C}_{AB} = 0$ while the mutual information is nonzero, then the extracted correlation by the detector is classical correlation or nondistillable entanglement.

\section{Results}\label{sec:results}

In this section, we compute the concurrence and the mutual information, given by Eqs.\eqref{eq: concurrence} and \eqref{eq: mutual information}, respectively, for the two qubtis in the scenario described by Fig. \ref{fig: entangled atoms}. The state of motion of both qubits is imprinted on the Wightman functions, given by Eqs.\eqref{eq: inertial wigthman}-\eqref{eq: rotating wigthman}, which allows us to construct $\mathcal{L}_{ij}$ and $\mathcal{M}$, the transition probabilities and `non-local' term, given by Eqs.\eqref{eq: transition probability} and \eqref{eq: non-local term}.

\subsection{Transition probabilities}

Let us start by considering the transition probability which are given by $\mathcal{L}_{ii}$ for $i = A~ \text{or}~ B$. We consider $\mathcal{L}_{ii}/\lambda^2$ and write the parameters in units of $\sigma$, which makes the transition probability a function of the three variables: $R/\sigma$, $E \sigma$ and $\Omega \sigma$.

\begin{figure}[ht!]
\centering
\includegraphics[scale=0.35]{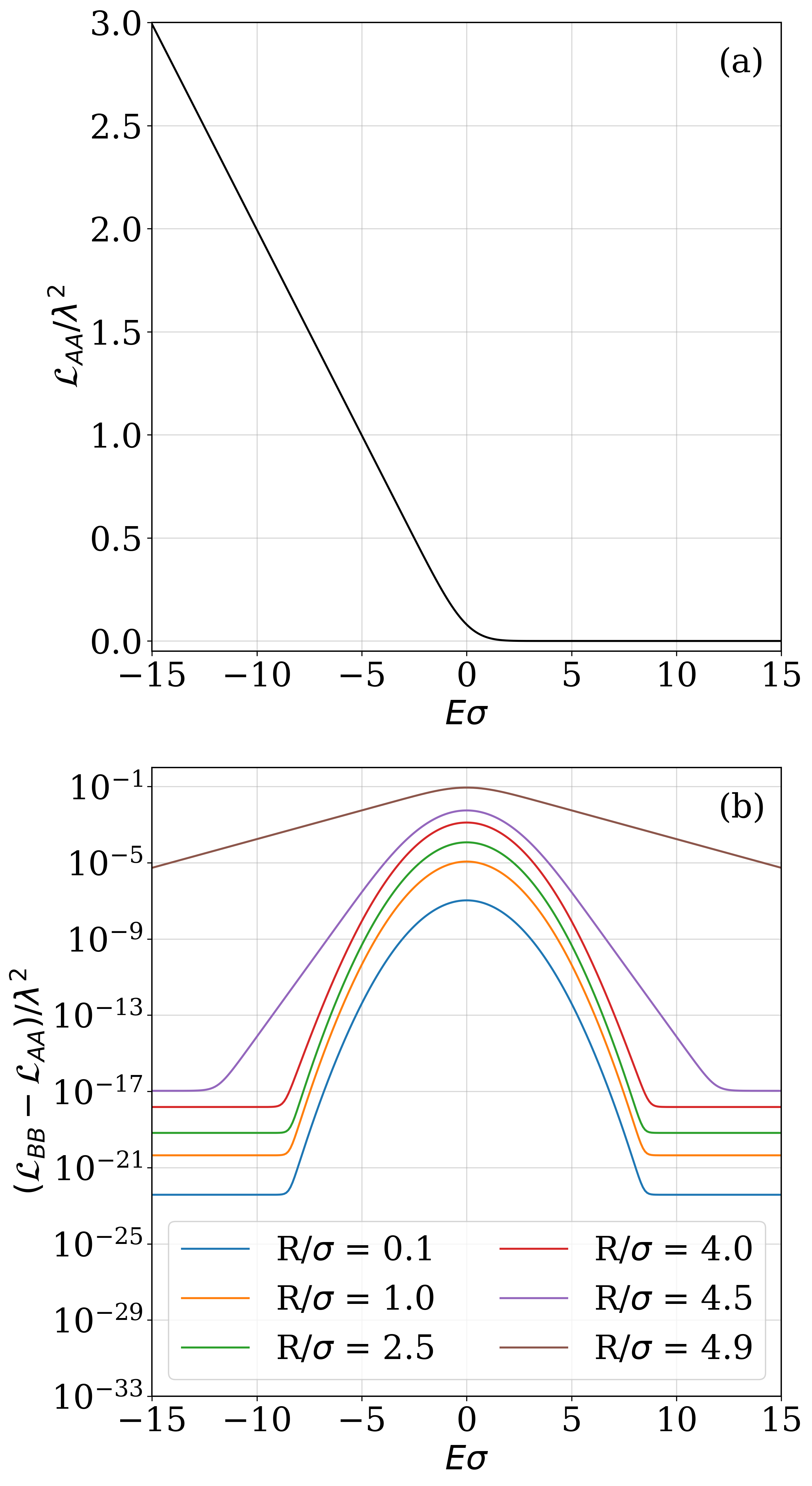}
\caption{(a) Transition probability $\mathcal{L}_{AA}/\lambda^2$ as a function of the energy gap $E\sigma$. (b) The term $( \mathcal{L}_{BB} - \mathcal{L}_{AA})/\lambda^2$ as a function of $E\sigma$ with $\Omega \sigma = 0.2$. Different values of $R/\sigma$ are represented by the different colors.
}
\label{fig:trans_prob}
\end{figure}

Fig. \ref{fig:trans_prob} shows the transition probability $\mathcal{L}_{ii}/\lambda^2$ as a function of the energy gap of the qubit multiplied by the time of interaction, $E\sigma$, for fixed values of radius $R/\sigma$ and the angular momentum is set as $\Omega \sigma = 5.0$. In Fig. \ref{fig:trans_prob}\textcolor{blue}{(a)}, we have the usual transition probability for an inertial trajectory, $\mathcal{L}_{11}/\lambda^2$, as shown in several works \cite{martin2016spacetime,tjoa2023}, while in Fig. \ref{fig:trans_prob}\textcolor{blue}{(b)}, we have the transition probability for a circular trajectory, $\mathcal{L}_{22}/\lambda^2$, which is consistent with the results in \cite{letaw80,takagi86}, with a subtle difference: in this work we use the previously defined switching function. We find that $\mathcal{L}_{22}/\lambda^2$ increases with the linear velocity. As $\mathcal{L}_{11}$ does not change with $R/\sigma$ and/or $\Omega \sigma$, we subtract this quantity from $\mathcal{L}_{22}$ in Fig. \ref{fig:trans_prob}\textcolor{blue}{(b)} to better analyse how variations in these parameters impact the curves. We have fixed $\Omega \sigma = 5.0$ and each color represent a value of $R/\sigma$, from $R/\sigma = 0.1$ to $R/\sigma = 4.9$. This quantity, $\mathcal{L}_{22} - \mathcal{L}_{11}$, is even having its peak at $E\sigma = 0$; both the width and the magnitude increase with $R/\sigma$.

\begin{figure*}[htp!]
\centering
\includegraphics[scale=0.35]{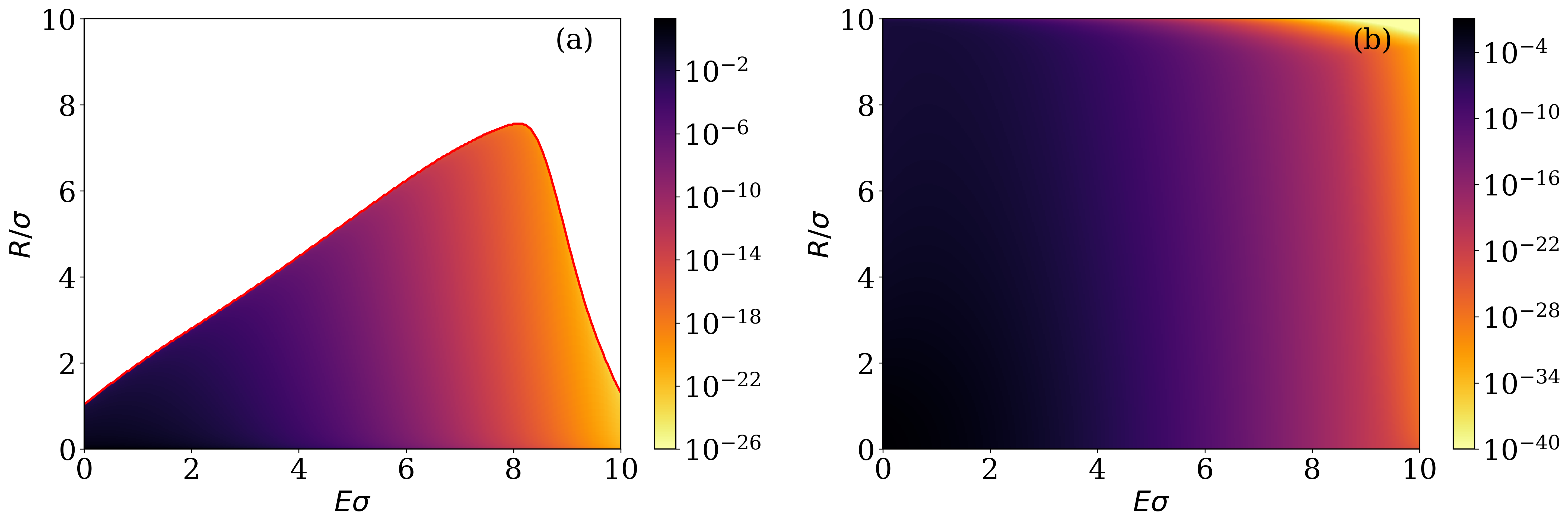}
\caption{(a) Concurrence $C_{AB}/\lambda^2$ with $\Omega \sigma = 0.1$ where the red line represent where $|\mathcal{M}| = \sqrt{\mathcal{L}_{AA} \mathcal{L}_{BB}}$ and the white region where $C_{AB} = 0$. (b) Mutual information $\mathcal{I}_{AB}/\lambda^2$ with $\Omega \sigma = 0.1$.}
\label{fig:conc_mInf}
\end{figure*}

\subsection{Concurrence and mutual information}

Next, we consider the concurrence $C_{AB}/\lambda^2$ and the mutual information $\mathcal{I}_{AB}/\lambda^2$ as functions of $R/\sigma$ and $E\sigma$ in Fig. \ref{fig:conc_mInf}. In these plots, we fix $\Omega \sigma = 0.1$. 
Both the concurrence and the mutual information decrease as $E\sigma$ increases. 
On the other hand, as $R/\sigma$ grows, both quantities remain approximately constant over a wide range of values. 
This behavior persists until, in the case of the concurrence, which can be seen in Fig. \ref{fig:conc_mInf}\textcolor{blue}{(a)}, the term $\sqrt{\mathcal{L}_{AA}\mathcal{L}_{BB}}$ becomes larger than $|\mathcal{M}|$, causing the concurrence to abruptly vanish.

\begin{figure}[ht!]
\centering
\includegraphics[scale=0.35]{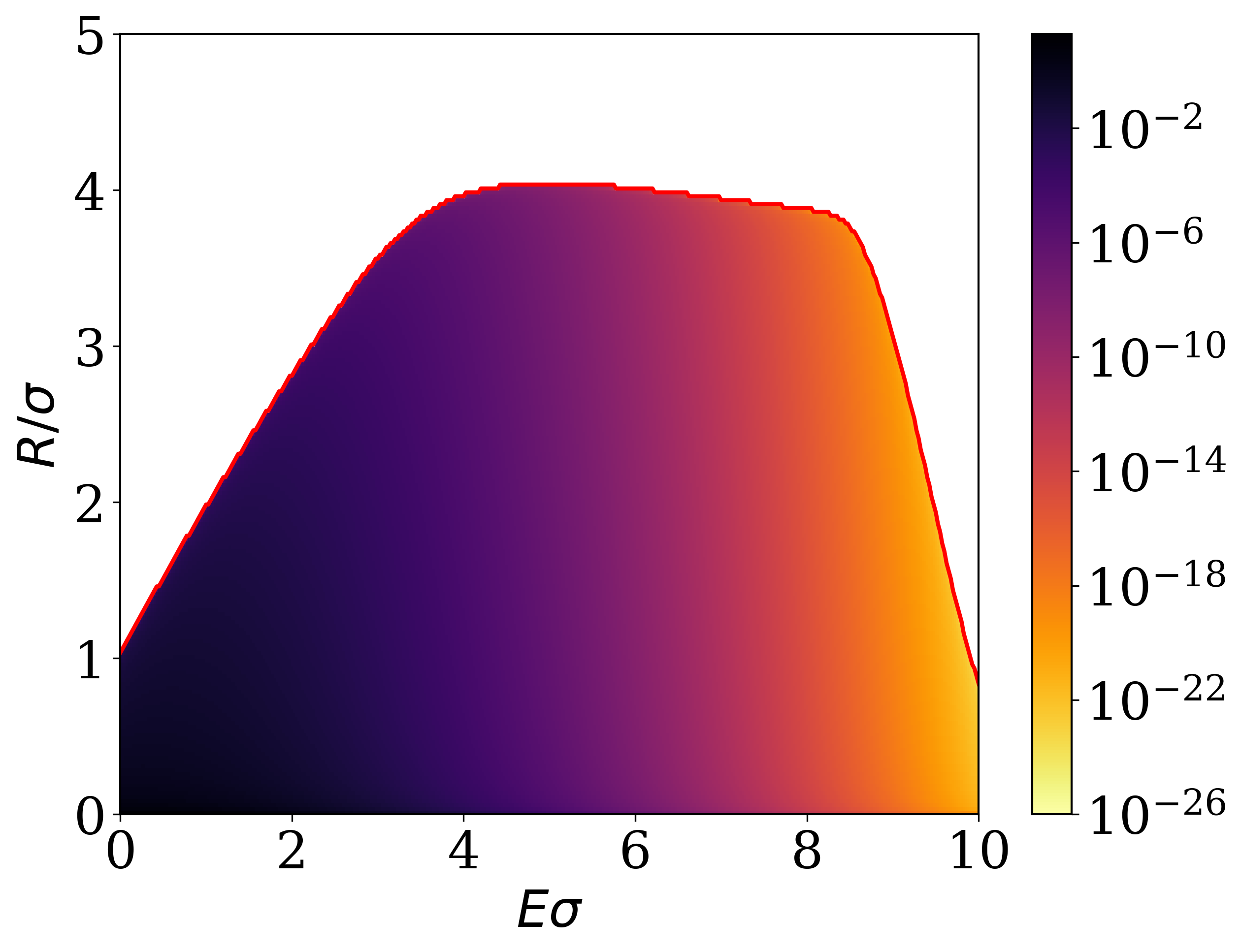}
\caption{Concurrence $C_{AB}/\lambda^2$ with $\Omega \sigma = 0.2$ where the red line represent where $|\mathcal{M}| = \sqrt{\mathcal{L}_{AA} \mathcal{L}_{BB}}$.}
\label{fig:concurence2}
\end{figure}

The concurrence cans persist up to relatively large values of $R/\sigma$ depending on $E\sigma$. For $\Omega \sigma = 0.1$, it is still possible to obtain $C_{AB} \neq 0$ for $R/\sigma \approx 7.5$. 
Moreover, as $E\sigma$ increases, the minimum value of the concurrence grows approximately linearly with $R/\sigma$ up to about $E\sigma \approx 8.2$, beyond which it begins to decrease rapidly. As we increase the angular velocity $\Omega \sigma$ (see Fig. \ref{fig:concurence2}), this behavior changes. The approximately linear growth persists; however, instead of extending up to $E\sigma \approx 8.2$, it now holds only until $E\sigma \approx 3.8$. 
From this point on, the curve develops an almost flat behavior up to $E\sigma \approx 8.2$, after which the minimum value once again decreases rapidly.

In contrast, the mutual information exhibits this nearly constant behavior as $R/\sigma$ increases until we approach relativistic regimes, namely when $R\Omega \approx 1$. 
Close to this threshold, the mutual information decreases sharply, as shown in Fig.~\ref{fig:conc_mInf}\textcolor{blue}{(b)}.

\section{Conclusion}\label{sec:conclusion}

In this work, we investigated entanglement harvesting between two qubits interacting with a massless scalar field. We assume that one of the qubits is inertial and the other undergoes uniform circular motion. By explicitly computing the two-point Wightman functions associated with inertial and rotating trajectories, we study trajectory-dependent entanglement generation.

Our results show that circular motion enhances the transition probability  of the rotating qubit ($\mathcal{L}_{BB}$), as the effect intensifies as the linear velocity increases. In particular, the difference between the rotating and inertial transition probabilities grows both in magnitude and width around $E\sigma = 0$ as $R/\sigma$ increases, revealing the strong sensitivity of excitation processes to rotational motion.

With respect to correlations, we have obtained that both concurrence and mutual information decrease as the energy gap $E\sigma$ increases. However, for fixed angular velocity, both quantities remain approximately constant over a wide range of radii $R/\sigma$, indicating a certain robustness of harvested correlations against changes in orbital radius.

A key feature of our results is the abrupt vanishing of concurrence when the term $\sqrt{\mathcal{L}_{AA}\mathcal{L}_{BB}}$ surpasses the `non-local' contribution $|\mathcal{M}|$. This behavior highlights the delicate balance between excitations and `non-local' correlations in entanglement harvesting protocol. By increasing the angular velocity modifies this balance significantly: for the maximum values of $R/\sigma$ that can generate entanglement, we observe an increase in the $E \sigma$ interval allowed, which can be viewed by comparison of Fig. \ref{fig:concurence2} with Fig. \ref{fig:conc_mInf}\textcolor{blue}{(a)}. This happens before a rapid degradation starts. Otherwise, the mutual information displays greater robustness, remaining nearly constant with increasing $R/\sigma$ until the onset of relativistic effects, characterized by $R\Omega \approx 1$. Close to this threshold, total correlations decrease sharply, signaling that extreme rotational motion strongly suppresses the ability of the detectors to extract correlations from the field.

Overall, our analysis demonstrates that circular motion introduces nontrivial kinematic effects in entanglement harvesting. While non-relativistic rotational motion allows significant extraction of quantum and classical correlations, relativistic regimes tend to degrade them. These results contribute to a deeper understanding of how non-inertial motion influences quantum correlations in relativistic quantum information settings and may be relevant for future studies involving rotating systems and relativistic quantum technologies.


\begin{acknowledgments}
This work was supported by Conselho Nacional de Desenvolvimento Cient\'{\i}fico e Tecnol\'{o}gico - CNPq, 303436/2015-8 (NFS) and FAPERJ PhD merit fellowship - FAPERJ Nota 10, 203.709/2025 (FS).
\end{acknowledgments}

\bibliography{IandR}

\newpage
\onecolumngrid
\appendix

\section{Explicit computation of $\mathcal{L}_{ij}$}\label{app: A}

Let us start by evaluating the term from Eq. \eqref{eq: transition probability} using the coordinates given by Eq. \eqref{eq: new variables}. The positive frequency Wightman function are given by Eqs. \eqref{eq: inertial wigthman}-\eqref{eq: rotating wigthman}. In the case where $i = j = 1$, we have
\begin{equation}
\mathcal{L}_{11} = \frac{\lambda^2}{4\pi^2}\lim_{\Delta x \to 0^+}\int_{-\infty}^{\infty}d\eta ~e^{-\frac{2\eta^2}{\sigma^2}} \int_{-\infty}^{\infty}d\xi ~e^{-\frac{\xi^2}{2\sigma^2} - i E \xi} \int_{0}^{\infty} dk ~e^{-i k \xi} \frac{\sin k \Delta x}{\Delta x}. \nonumber
\end{equation}
In the above equation, we integrate both Gaussian integrals and obtain
\begin{equation}
\mathcal{L}_{11} = \frac{\lambda^2 \sigma^2}{4\pi}\lim_{\Delta x \to 0^+} \int_0^\infty dk ~e^{-\frac{1}{2}(E+k)^2\sigma^2} \frac{\sin k \Delta x}{\Delta x} \equiv \frac{\lambda^2}{4\pi} \left[ \exp\left(-\frac{E^2 \sigma^2}{2} \right) - \sqrt{\frac{\pi}{2}} E \sigma ~ \text{Erfc}\left(\frac{E \sigma}{\sqrt{2}} \right) \right],
\end{equation}
which is the same expression as presented in Eq. \eqref{eq: transition probability 11}.

For $i = 1$ and $j = 2$, we use the following coordinate change
 \begin{equation}\label{eq: tau and tau prime relation}
\tau - \gamma \tau' = -(\gamma - 1)\eta + \frac{1}{2}(\gamma +1)\xi \equiv -\gamma_- \eta + \frac{1}{2}\gamma_+ \xi,
\end{equation}
where $\gamma_\pm = \gamma \pm 1$. Therefore, we have
\begin{equation}
\mathcal{L}_{12} = \frac{\lambda^2}{4\pi^2}\int_0^\infty dk ~\frac{\sin k R}{R} \int_{-\infty}^{\infty}d\eta ~e^{- \frac{2\eta^2}{\sigma^2}+ i k \gamma_- \eta} \int_{-\infty}^{\infty}d\xi ~e^{- \frac{\xi^2}{2\sigma^2} - i E \xi - i \frac{k \gamma_+}{2}\xi}. \nonumber
\end{equation}
By using the results of the integrals
\begin{equation}
\int_{-\infty}^{\infty}d\eta ~e^{- \frac{2\eta^2}{\sigma^2}+ i k \gamma_- \eta} = \sqrt{\frac{\pi}{2}} \sigma ~e^{-\frac{\gamma^2_-}{8} k^2 \sigma^2}, \quad \int_{-\infty}^{\infty}d\xi ~e^{- \frac{\xi^2}{2\sigma^2} - i E \xi - i \frac{k \gamma_+}{2}\xi} = \sqrt{2\pi} \sigma ~e^{-\frac{1}{8}(\gamma_+ k + 2E)^2 \sigma^2}, \nonumber
\end{equation}
$\mathcal{L}_{12}$ can be rearranged as
\begin{equation}
\mathcal{L}_{12} = \frac{\lambda^2 \sigma^2}{4\pi}e^{-\frac{1}{2}E^2 \sigma^2}\int_0^\infty dk ~\frac{\sin k R}{R} ~e^{-\frac{1}{8}(\Sigma^2 k^2 + 4\gamma_+ E k) \sigma^2}, \nonumber
\end{equation}
where $\Sigma^2 = \gamma^2_+ + \gamma^2_-$. By solving the above integral one can obtain
\begin{equation}
\mathcal{L}_{12} = -\frac{i}{4\sqrt{2\pi}} \frac{\lambda^2 \sigma}{\Sigma R} e^{-\frac{1}{2}E^2 \sigma^2}\left[ e^{-j_+^2} \text{Erfc}(j_-) - e^{-j_-^2} \text{Erfc}(j_+) \right],
\end{equation}
where $j_\pm = \frac{1}{\sqrt{2}\Sigma} \left( \gamma_+ E\sigma \pm 2i R/\sigma \right)$. The previous equation is exactly the same as the one given by Eq. \eqref{eq: transition probability 12}.

For the case where $i = j = 2$ we have:
\begin{equation}\label{eq: L22 v0}
\mathcal{L}_{22} = \frac{\lambda^2}{4\pi^2} \int_{-\infty}^{\infty} d\eta~e^{-\frac{2\eta^2}{\sigma^2}} \int_{0}^{\infty} dk \int_{-\infty}^{\infty} d\xi ~e^{-\frac{\xi^2}{2\sigma^2} - iE \xi - i \gamma k \xi} \frac{\sin 2k R \left|\sin\frac{\gamma \Omega}{2} \xi \right| }{2R \left|\sin\frac{\gamma \Omega}{2} \xi \right|},
\end{equation}
due the complexity of the last term, we will need to integrate over $k$ first and obtain a `distribution'. It is well-known that the positive Wightman function cane be written as
\begin{equation}
    W(x;x')=\frac{1}{4\pi^2} \int_{0}^{\infty} d\omega ~e^{-i\omega(t - t')} \frac{\sin \omega|\mathbf{x} - \mathbf{x}'|}{|\mathbf{x} - \mathbf{x}'|} \equiv -\frac{1}{4\pi^2} \frac{1}{(t-t')^2 - |\mathbf{x} - \mathbf{x}'|^2},
\end{equation}
which, using Eq. \eqref{eq: rotating observer} for $t$ and $\mathbf{x}$, can be written as 
\begin{equation}
    W(x_2;x'_2) = -\frac{1}{4\pi^2} \frac{1}{\gamma^2\xi^2 - 4 R^2 \sin^2\left(\frac{\gamma \Omega}{2}\xi\right)}. \nonumber
\end{equation}

The denominator of the previous expression possesses only one root ($\xi = 0$). In order to remove this pole, we can sum and subtract $1/\xi^2$, obtaining
\begin{equation}
    W(x_2;x'_2) = W(x_1;x'_1) + \frac{1}{4\pi^2}\left(\frac{1}{\xi^2} - \frac{1}{\gamma^2\xi^2 - 4 R^2 \sin^2\left(\frac{\gamma \Omega}{2}\xi\right)} \right), \nonumber
\end{equation}
where, combining the last term, can be written as
\begin{equation}\label{eq: simplified rotating wightman function}
    W(x_2;x'_2) = W(x_1;x'_1) - \frac{1}{4\pi^2} \frac{(1-\gamma^2) \xi^2 + 4R^2 \sin^2\left(\frac{\gamma \Omega}{2}\xi \right)}{\xi^2\left[\gamma^2\xi^2 - 4 R^2 \sin^2\left(\frac{\gamma \Omega}{2}\xi\right) \right]}.
\end{equation}
Replacing Eq. \eqref{eq: simplified rotating wightman function} in Eq. \eqref{eq: L22 v0}, we obtain
\begin{equation}
    \mathcal{L}_{22} = \mathcal{L}_{11} - \frac{\lambda^2}{4\pi^2} \int_{-\infty}^{\infty} d\eta~e^{-\frac{2\eta^2}{\sigma^2}} \int_{-\infty}^{\infty} d\xi ~e^{-\frac{\xi^2}{2\sigma^2} - iE \xi} \frac{(1-\gamma^2) \xi^2 + 4R^2 \sin^2\left(\frac{\gamma \Omega}{2}\xi \right)}{\xi^2\left[\gamma^2\xi^2 - 4 R^2 \sin^2\left(\frac{\gamma \Omega}{2}\xi\right) \right]}, \nonumber
\end{equation}
the first integral is just a Gaussian one as we already have computed before. For simplicity, we change the variable by defining
\begin{equation}
    v = \Omega R, ~z = \frac{\gamma \Omega}{2}\xi, \nonumber
\end{equation}
which lead us to
\begin{equation}
    \mathcal{L}_{22} = \mathcal{L}_{11} - \frac{\lambda^2 \Omega\sigma}{8\pi^2\gamma} \sqrt{\frac{\pi}{2}}  \int_{-\infty}^{\infty} dz ~e^{-2\left[ \left(\frac{z}{\gamma \Omega\sigma}\right)^2 + i\frac{E z}{\gamma \Omega}\right]} \frac{(1-\gamma^2) z^2 + \gamma^2 v^2 \sin^2 z}{z^2\left(z^2 - v^2 \sin^2 z \right)}. \nonumber
\end{equation}
Since $1 - \gamma^2 = -\gamma^2v^2$, and with the symmetrical interval of integration, we obtain
\begin{equation}
    \mathcal{L}_{22} = \mathcal{L}_{11} +  \frac{\lambda^2}{4\pi^2} \sqrt{\frac{\pi}{2}}~ \gamma v^2\Omega\sigma \int_{0}^{\infty} dz ~e^{-2\left(\frac{z}{\gamma \Omega\sigma}\right)^2} \cos\left(\frac{2E z}{\gamma \Omega} \right) \frac{z^2 -\sin^2 z}{z^2\left(z^2 - v^2 \sin^2 z \right)},
\end{equation}
which is exactly the same expression as the one given by Eq. \eqref{eq: transition probability 22}.

\section{Explicit computation of the `non-local' term $\mathcal{M}$}
\label{app:B}

In this appendix we present the explicit evaluation of the `non-local' contribution
$\mathcal{M}$ defined in Eq.~\eqref{eq: non-local term}. We make use of the change of variables $(\tau,\tau') \mapsto (\eta,\xi)$ introduced in Eq.~\eqref{eq: new variables}. This change modifies the integration domain as the following
\begin{equation}
    \tau' \leq \frac{\tau}{\gamma},
    \quad
    \tau \leq \gamma \tau'. \nonumber
\end{equation}
In terms of $(\eta,\xi)$, this implies
\begin{align}
    \gamma\!\left(\eta - \frac{\xi}{2}\right) \leq \eta + \frac{\xi}{2}
    &\quad \Rightarrow \quad
    \frac{2\gamma_-}{\gamma_+}\,\eta \leq \xi,
    \\
    \eta + \frac{\xi}{2} \leq \gamma\!\left(\eta - \frac{\xi}{2}\right)
    &\quad \Rightarrow \quad
    \xi \leq \frac{2\gamma_-}{\gamma_+}\,\eta, \nonumber
\end{align}
where we have defined $\gamma_\pm = \gamma \pm 1$. Using the positive frequency Wightman function given by Eq.~\eqref{eq: crossed wigthman 12}, the identity $W(x'_2;x_1) = W(x_1;x'_2)^{*}$ and the relation between $\tau$ and $\tau'$ from Eq.~\eqref{eq: tau and tau prime relation}, we obtain
\begin{align}
    \mathcal{M}
    ={}&
    -\frac{\lambda^2}{4\pi^2}
    \int_{0}^{\infty}\! dk\, \frac{\sin(kR)}{R}
    \int_{-\infty}^{\infty}\! d\eta\,
    e^{-\frac{2\eta^2}{\sigma^2} + i(2E + k\gamma_-)\eta}
    \int_{\frac{2\gamma_-}{\gamma_+}\eta}^{\infty}\! d\xi\,
    e^{-\frac{\xi^2}{2\sigma^2} - i\frac{k\gamma_+}{2}\xi}
    \nonumber\\
    &-
    \frac{\lambda^2}{4\pi^2}
    \int_{0}^{\infty}\! dk\, \frac{\sin(kR)}{R}
    \int_{-\infty}^{\infty}\! d\eta\,
    e^{-\frac{2\eta^2}{\sigma^2} + i(2E - k\gamma_-)\eta}
    \int_{-\infty}^{\frac{2\gamma_-}{\gamma_+}\eta}\! d\xi\,
    e^{-\frac{\xi^2}{2\sigma^2} + i\frac{k\gamma_+}{2}\xi}.
\end{align}
The integrals over $\xi$ can be performed analytically, which yields
\begin{align}
    \mathcal{M}
    ={}&
    -\frac{\lambda^2 \sigma}{4\pi^2}\sqrt{\frac{\pi}{2}}
    \int_{0}^{\infty}\! dk\, \frac{\sin(kR)}{R}
    e^{-\frac{1}{8}\gamma_+^2 \sigma^2 k^2}
    \int_{-\infty}^{\infty}\! d\eta\,
    e^{-\frac{2\eta^2}{\sigma^2} + i(2E + k\gamma_-)\eta}
    \Bigg[
        1 - \mathrm{Erf}\!\left(
        \frac{\sqrt{2}\gamma_-}{\gamma_+\sigma}\eta
        + i\frac{\gamma_+\sigma k}{2\sqrt{2}}
        \right)
    \Bigg]
    \nonumber\\
    &-
    \frac{\lambda^2 \sigma}{4\pi^2}\sqrt{\frac{\pi}{2}}
    \int_{0}^{\infty}\! dk\, \frac{\sin(kR)}{R}
    e^{-\frac{1}{8}\gamma_+^2 \sigma^2 k^2}
    \int_{-\infty}^{\infty}\! d\eta\,
    e^{-\frac{2\eta^2}{\sigma^2} + i(2E - k\gamma_-)\eta}
    \Bigg[
        1 + \mathrm{Erf}\!\left(
        \frac{\sqrt{2}\gamma_-}{\gamma_+\sigma}\eta
        - i\frac{\gamma_+\sigma k}{2\sqrt{2}}
        \right)
    \Bigg]. \nonumber
\end{align}

After a few manipulations on the above equation, one can write
\begin{equation}
    \mathcal{M}
    =
    -\frac{\lambda^2 \sigma}{4\pi^2}\sqrt{\frac{\pi}{2}}
    \int_{0}^{\infty}\! dk\, \frac{\sin(kR)}{R}
    e^{-\frac{1}{8}\gamma_+^2 \sigma^2 k^2}
    \int_{-\infty}^{\infty}\! d\eta\,
    e^{-\frac{2\eta^2}{\sigma^2} + 2iE\eta}
    \Big[2 I_0(\eta,k) - I_1^{*}(\eta,k) + I_1(\eta,k)\Big],
\end{equation}
where
\begin{align}
    I_0(\eta,k) &= \cos(k\gamma_- \eta), \\
    I_1(\eta,k) &=
    e^{-ik\gamma_- \eta}
    \mathrm{Erf}\!\left(
    \frac{\sqrt{2}\gamma_-}{\gamma_+\sigma}\eta
    - i\frac{\gamma_+\sigma k}{2\sqrt{2}}
    \right).
\end{align}
The contribution coming from the integration of $I_0$ is
\begin{eqnarray}
    \!\int_{-\infty}^{\infty}\! d\eta\,
    e^{-\frac{2\eta^2}{\sigma^2} + 2iE\eta} I_0(\eta,k)
    &=&
    \frac{1}{2}\sqrt{\frac{\pi}{2}}\,\sigma
    \left[
        e^{-\frac{1}{8}(2E + k\gamma_-)^2 \sigma^2}
        +
        e^{-\frac{1}{8}(2E - k\gamma_-)^2 \sigma^2}
    \right]  \nonumber \\
    &=&
    \sqrt{\frac{\pi}{2}}\,\sigma
    e^{-\frac{1}{2}E^2 \sigma^2 - \frac{1}{8}\gamma_-^2 k^2 \sigma^2} \cosh \left( \frac{\gamma_-}{2} E k \sigma^2  \right), \nonumber
\end{eqnarray}
 and the contribution from $I_1$ and $I^*_1$ are
\begin{eqnarray}
    \int_{-\infty}^{\infty}\! d\eta\,
    e^{-\frac{2\eta^2}{\sigma^2} + 2iE\eta} I_1(\eta,k)
    &=&
    +\sqrt{\frac{\pi}{2}}\,\sigma\,
    e^{-\frac{1}{8}(2E - k\gamma_-)^2 \sigma^2}
    \mathrm{Erf}\!\left[
        \frac{i\sigma}{\sqrt{2}}\left( \frac{k \Sigma}{2} - \frac{E \gamma_-}{\Sigma} \right)
    \right], \nonumber \\
    \int_{-\infty}^{\infty}\! d\eta\,
    e^{-\frac{2\eta^2}{\sigma^2} + 2iE\eta} I^*_1(\eta,k)
    &=&
    -\sqrt{\frac{\pi}{2}}\,\sigma\,
    e^{-\frac{1}{8}(2E + k\gamma_-)^2 \sigma^2}
    \mathrm{Erf}\!\left[
        \frac{i\sigma}{\sqrt{2}}\left( \frac{k \Sigma}{2} + \frac{E \gamma_-}{\Sigma} \right)
    \right], \nonumber
\end{eqnarray}
where $\Sigma^2 = \gamma_-^2 + \gamma_+^2$. We then write
\begin{eqnarray}
    \mathcal{M}
    &=&
    -\frac{\lambda^2 \sigma^2}{8\pi}
    e^{-\frac{1}{2}E^2\sigma^2}
    \int_{0}^{\infty}\! dk\, \frac{\sin(kR)}{R}
    e^{-\frac{1}{8}\Sigma^2 k^2 \sigma^2} \\
    &&\times
    \left\{
        2\cosh \left( \frac{\gamma_-}{2} E k \sigma^2  \right)
        + e^{-\frac{\gamma_-}{2}E k \sigma^2} \mathrm{Erf}\!\left[
        \frac{i\sigma}{\sqrt{2}}\left( \frac{k \Sigma}{2} + \frac{E \gamma_-}{\Sigma} \right)
    \right] + e^{\frac{\gamma_-}{2}E k \sigma^2} \mathrm{Erf}\!\left[
        \frac{i\sigma}{\sqrt{2}}\left( \frac{k \Sigma}{2} - \frac{E \gamma_-}{\Sigma} \right)
    \right]
    \right\}, \nonumber
\end{eqnarray}
which is exactly the same expression as the one given by Eq. \eqref{eq: non-local term M}.

\end{document}